# Magnetic Entropy in a Non-Collinear Weak Ferromagnetic YCrO$_3$

Brajesh Tiwari[1, a)], Ambesh Dixit[2, a)] and M. S. Ramachandra Rao[3]

[1]Department of Physics, Institute of Infrastructure Technology Research and Management Ahmedabad 380026, India

[2]Department of Physics & Center for Solar Energy, Indian Institute of Technology Jodhpur, 342037, India

[3]Department of Physics, Indian Institute of Technology Madras, 600036, India

a)Corresponding author: brajeshtiwari@iitram.ac.in

**Abstract:**

We carried out temperature and field dependent magnetic measurements to understand the evolution of magnetic non-collinearity near antiferromagnetic phase in conjunction with the evolution of magnetic entropy near phase transition. We observed the maximum change in entropy just before magnetic ordering of $Cr^{3+}$ in YCrO$_3$ with the maximum change in magnetic entropy of ~ -0.38 Jkg$^{-1}$K$^{-1}$ at 8 Tesla external field. The data is linear in higher fields (3 T – 8 T), whereas it showed deviations in the lower field region. The maximum entropy change fits well with mean field approximation at higher fields, while the observed deviation in lower field substantiates the onset of weak ferromagnetism in YCrO$_3$.

# INTRODUCTION

$ACrO_3$ (A = rare earth elements) systems exhibit the stable G-type antiferromagnetic (AFM) ground state among other possible AFM structures. In addition, the non-collinear spin arrangement of $Cr^{3+}$ magnetic ions in these G-type AFM materials, gives rise to the weak ferromagnetism. The ratio c/a is very crucial for orthorhombic distortion in $ACrO_3$ which is one of the main sources for weak ferromagnetism via Dzialoshinskii–Moriya (DM) interaction $D_{ij} = \sum_{ij} D_{ij} \cdot (S_i \times S_j)$, where $S_i$ is the spint vector at $i^{th}$ atomic site. DM interaction arises from spin-orbit coupling of hopping electrons in an inversion asymmetrical crystal field. Depending on sign, the symmetry properties and the value of $D_{ij}$ vector, a directional non-collinear magnetic structure is possible in AFM materials. Among different orthochromites, $YCrO_3$ is an interesting material, which exhibits weak ferromagnetic behavior at or below its' Neel temperature 140 K [1]. The onset of weak ferromagnetism in this system is attributed to the canted G-type antiferromagnetic structure, where adjacent chromium octahedra (Cr-O1-Cr) are tilted at an angle of ~ 147º [1, 2]. The schematic crystallographic structure and in plane corner sharing of $CrO_6$ octahedra with Cr-O2-Cr angle about 149º are shown in Fig 1. This tilt may lead to unusual magnetophysical properties such as spin-lattice coupling, dielectric and ferroelectric response with magnetic field in conjunction with weak ferromagnetism in $YCrO_3$. Thus, it is important to understand the evolution of magnetic ordering near transition temperature in $YCrO_3$ and any associated magnetophysical properties. The associated change in the magnetic ordering may give rise to change in magnetic entropy. The evolution of magnetic entropy as a function of temperature will exhibit drastic changes near phase transition. This magnetic phase transition together with weak ferromagnetism in $YCrO_3$ will exhibit the respective change in magnetic entropy. We will discuss in detail about the observed degree of non-collinearity in $YCrO_3$ bulk samples.

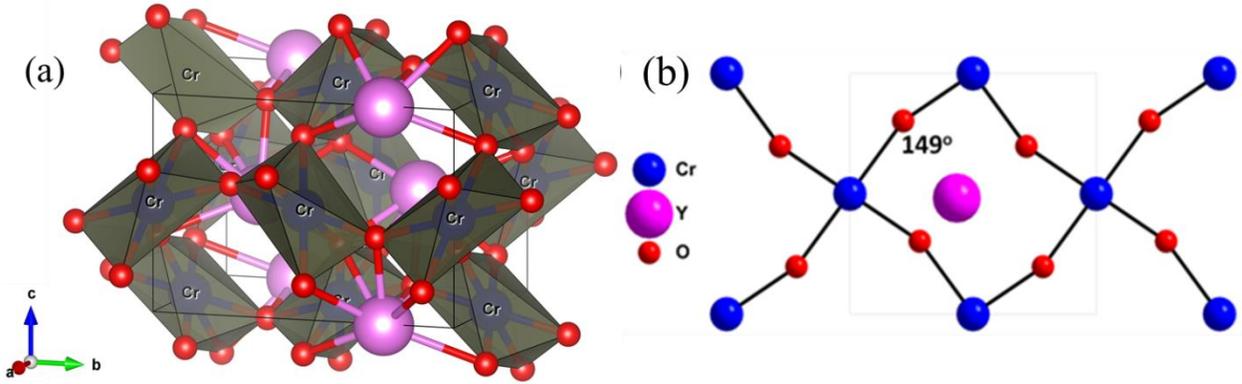

**FIGURE 1.** (a) Crystal structure of $YCrO_3$ at room temperature in Pnma space group where Cr occupies the center of octahedral. (b) in plane corner sharing of octahedra $CrO6$ with Cr-O2-Cr angle about 149°.

**EXPERIMENTAL DETAILS**

Bulk $YCrO_3$ was synthesized by conventional solid-state reaction method by mixing $Y_2O_3$ and $Cr_2O_3$ in stoichiometric ratio. Thoroughly mixed powder was first calcinated at 600 °C for 12 hours and then at 900 °C for 12 hours before final annealing at 1300 °C for 24 hours. The detailed of synthesis method and other characterizations can be found in our previous work [1]. The magnetization isotherms were measured using a vibrating sample magnetometer, an attachment in PPMS (Model 6000, Quantum Design, USA).

**RESULTS AND DISCUSSION**

The synthesized compound $YCrO_3$ is showing orthorhombic crystal structure with Pnma space group, Fig 1 (a). The structural details are published elsewhere [1]. This is a distorted $GdFeO_3$-type perovskite structure with out of plane Cr-O1-Cr octahedra bond angles (~147°) and in plane Cr-O2-Cr octahedral bond angle (~ 149°), deviating from undistorted 180°, as shown schematically in Fig 1 (b). The spin magnetic moments follow the adjacent octahedral tilt due to spin-orbit coupling though small which in turn lead to canted G-type antiferromagnetic ordering below 140 K. This canted spin arrangement in G-type antiferromagnet is the main source of weak ferromagnetism due the DM interaction and consistent with the literature [1,9,10].

The magnetic properties of YCrO$_3$ are investigated in depth, showing the onset of hysteresis in low field magnetic hysteresis measurements below the anti-ferromagnetic transition temperature ~ 140K [8, 10]. On the other hand, temperature dependent magnetic measurements exhibit the antiferromagetnic ordering at 140 K [1,2, 8, 10]. We carried out temperature and field dependent magnetization measurements to understand the onset of magnetic entropy and correlation with observed non-collinearity in YCrO$_3$ system. The recorded magnetic isotherms at different temperatures are shown in Fig 2 (a), exhibiting the nonlinear region at or below 3 T magnetic field and the nature becomes linear for higher fields. The noticed low field nonlinearity in magnetic data can be understood in terms of non-collinear magnetic structure, giving rise to the weak ferromagnetism in conjunction with AFM structure. Thermodynamics relates the magnetic variables to entropy and temperature. This is due to the coupling of magnetic sublattice to the external field which contributes to change in magnetic entropy of magnetic materials. Isothermal magnetic entropy $S(T, H)$ can be estimated from magnetization, $M(T, H)$, data as a function of applied field ($H$) recorded at different temperature ($T$) especially near magnetic transitions. The thermodynamic Maxwell's relation $\left(\frac{\partial S(T,H)}{\partial H}\right)_T = \left(\frac{\partial M(T,H)}{\partial T}\right)_H$ will lead to magnetic entropy change from numerical integration as $\Delta S(T, \Delta H) = \mu_0 \int_{H_i}^{H_f} \left(\frac{\partial M}{\partial T}\right)_H dH$, where $\Delta H = H_f - H_i$ and usually $H_i$ is taken as zero magnetic fields. For magnetically ordered system entropy change is the maximum about transition temperature however for paramagnets it is significant near absolute temperature. The magnetic isotherms, Fig 2 (a), are used to estimate the change in magnetic entropy against temperature at different magnetic field. The computed magnetic entropy at different fields is shown in Fig 2 (b). We observed the evolution of magnetic entropy near magnetic phase transition at about 140 K, which reduces quickly at higher temperatures. The change in entropy is increasing with increasing magnetic field and the maximum entropy change of ~ -0.38 Jkg$^{-1}$K$^{-1}$ at 8 T field is recorded, Fig 2 (b). The amplitude of the maxima of change in entropy is reducing with lowering field, Fig 2(b). The evolution of magnetic entropy near magnetic phase transition in YCrO$_3$ is relatively larger than respective antiferromagnetic transitions, suggesting the impact of weak ferromagnetism near transition.

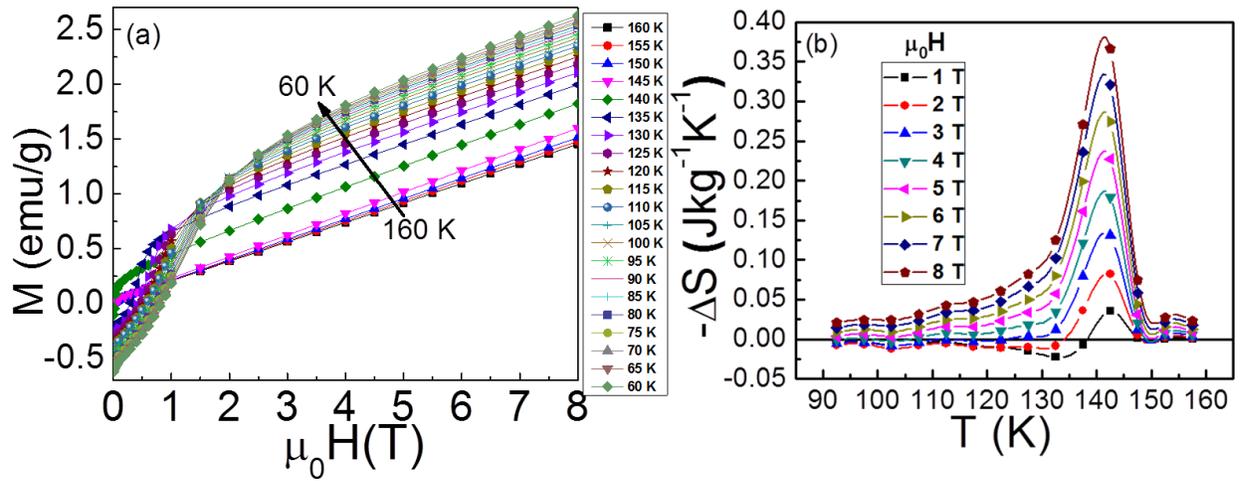

**FIGURE 2.** (a) Magnetization isotherms recorded at different temperatures in interval of 5 K. (b) Change in magnetic entropy as a function of temperature about the magnetic transition temperature.

## CONCLUSION

We demonstrated the evolution of magnetic entropy in YCrO$_3$ near antiferromagnetic ordering temperature ~ 140 K. The recorded values of the maximum change in entropy ~ -0.38 Jkg$^{-1}$K$^{-1}$ at 8 Tesla is relatively larger than expected for pure antiferromagnetic system, supporting the observation of weak ferromagnetism in YCrO$_3$ near the transition temperature. Further, weak ferromagnetism in YCrO$_3$ is substantiated by the observation of nonlinear magnetic behavior up to 3 T whereas magnetization becomes linear at and above 3 T in field dependent magnetic measurements.

## ACKNOWLEDGMENT:

Author Brajesh Tiwari acknowledges Professor Shiva Prasad for his technical discussion during the manuscript preparation and Ambesh Dixit highly acknowledges the financial assistance from UGC-DAE Consortium For Scientific Research, Gov. of India through project number CRS-M-221 for this work.**REFERENCES:**

1. B. Tiwari, M. K. Surendra, M. S. Ramachandra Rao, HoCrO$_3$ and YCrO$_3$: a comparative study, J. Phys.: Condens. Matter 25, 216004 (2013)

2. J. Lyubina, M. D. Kuzmin, K. Nenkov O. Gutfleisch, M. Richter, D. L. Schlagel, T. A. Lograsso, K. A. Gschneidner Jr, Magnetic field dependence of the maximum magnetic entropy change, Phys. Rev. B, 83, 012403 (2011)

3. M. P. Brown and K. Austin, Appl. Phys. Letters **85**, 2503–2504 (2004).

4. R. T. Wang, "Title of Chapter," in *Classic Physiques*, edited by R. B. Hamil (Publisher Name, Publisher City, 1999), pp. 212–213.

5. C. D. Smith and E. F. Jones, "Load-cycling in cubic press," in *Shock Compression of Condensed Matter-2001*, AIP Conference Proceedings 620, edited by M. D. Furnish *et al*. (American Institute of Physics, Melville, NY, 2002), pp. 651–654.

6. B. R. Jackson and T. Pitman, U.S. Patent No. 6,345,224 (8 July 2004)

7. D. L. Davids, "Recovery effects in binary aluminum alloys," Ph.D. thesis, Harvard University, 1998.

8. V. M. Judin and A. B. Sherman "Weak ferromagnetism of YCrO$_3$" Solid State Comm. **4**, 661-663 (1966)

9. Y. Sharma et.al "Phonons and magnetic excitation correlations in weak ferromagnetic YCrO$_3$" Journal of Applied Physics **115**, 183907 (2014)

10. C. Alvarez, M. P. Cruz, A. C. Duran, H. Montiel, R. Zamorano, Weak ferromagnetism in the magnetoelectric YCrO3 detected by microwave power absorption measurements, Solid State Communication, 150, 1597-1600 (2010)
6